\documentclass[prb, showpacs, twocolumn, superscriptaddress]{revtex4}
\usepackage{graphicx}
\usepackage{epstopdf}
\usepackage{amssymb}
\usepackage{subfigure}

\begin{document}
\title{The suppression of magnetism and the development of superconductivity within the collapsed tetragonal phase of Ca$_{0.67}$Sr$_{0.33}$Fe$_2$As$_2$ at high pressure}

\author{J. R. Jeffries}
\affiliation{Condensed Matter and Materials Division, Lawrence Livermore National Laboratory, Livermore, CA 94550, USA}
\author{N. P. Butch}
\affiliation{Condensed Matter and Materials Division, Lawrence Livermore National Laboratory, Livermore, CA 94550, USA}
\author{K. Kirshenbaum}
\affiliation{Center for Nanophysics and Advanced Materials, Department of Physics, University of Maryland, College Park, MD 20742, USA}
\author{S. R. Saha}
\affiliation{Center for Nanophysics and Advanced Materials, Department of Physics, University of Maryland, College Park, MD 20742, USA}
\author{S. T. Weir}
\affiliation{Condensed Matter and Materials Division, Lawrence Livermore National Laboratory, Livermore, CA 94550, USA}
\author{Y. K. Vohra}
\affiliation{Department of Physics, University of Alabama at Birmingham, Birmingham, Alabama 35294, USA}
\author{J. Paglione}
\affiliation{Center for Nanophysics and Advanced Materials, Department of Physics, University of Maryland, College Park, MD 20742, USA}

\date\today

\begin{abstract}
Structural and electronic characterization of (Ca$_{0.67}$Sr$_{0.33}$)Fe$_2$As$_2$ has been performed as a function of pressure up to 12 GPa using conventional and designer diamond anvil cells. The compound (Ca$_{0.67}$Sr$_{0.33}$)Fe$_2$As$_2$ behaves intermediate between its end members---CaFe$_2$As$_2$ and SrFe$_2$As$2$---displaying a suppression of magnetism and the onset of superconductivity. Like other members of the AEFe$_2$As$_2$ family, (Ca$_{0.67}$Sr$_{0.33}$)Fe$_2$As$_2$ undergoes a pressure-induced isostructural volume collapse, which we associate with the development of As-As bonding across the mirror plane of the structure. This collapsed tetragonal phase abruptly cuts off the magnetic state, giving rise to superconductivity with a maximum $T_c$=22.2 K. The maximum $T_c$ of the superconducting phase is not strongly correlated with any structural parameter, but its proximity to the abrupt suppression of magnetism as well as the volume collapse transition suggests that magnetic interactions and structural inhomogeneity may play a role in its development. The pressure-dependent evolution of the ordered states and crystal structures in (Ca,Sr)Fe$_2$As$_2$ provides an avenue to understand the generic behavior of the other members of the AEFe$_2$As$_2$ family.

\end{abstract}



\maketitle

\section{Introduction}
Since the first reports of superconductivity with a critical temperature $T_c$=26 K in fluorine-doped LaFeAsO,\cite{Kamihara2008} researchers have rapidly expanded the number of Fe-based superconductors,\cite{Ishida2009} raised the $T_c$ to about 55 K,\cite{Ren2008} and identified five different, but related, crystal structures in which these Fe-based superconductors crystallize.\cite{Paglione2010, Johnston2010} Like the cuprate superconductors,\cite{Lee2006} the different Fe-based superconductors display many common themes in both the electronic and structural properties: the presence of corrugated Fe-pnictogen or Fe-chalcogen layers within a tetragonal unit cell, and the occurrence of antiferromagnetic order in the undoped or ambient-pressure compounds.\cite{Lumsden2010} The ubiquity of these common elements makes these systems fertile playgrounds to explore the interplay between magnetism, structure, and superconductivity.

One of the archetypal Fe-based superconductor structures is the ``122'' structure: AEFe$_2$X$_2$ (ThCr$_2$Si$_2$-type), with AE an alkaline earth element (Ca, Sr, Ba) an alkali metal (K, Rb, Cs) or Eu and X a pnictogen element.\cite{Paglione2010, Johnston2010, Lumsden2010, Gooch2010, Jeevan2008} Variants of the 122 structure have been widely studied owing to the availability of a wide range of chemical substitutions on different crystallographic sites ({\it{e.g.}}, Co for Fe, K for Ba, P for As, etc.) as well as their tendency to form macroscopic, high-purity crystals. The parent compounds within the 122 systems are paramagnetic metals at room temperature, but at low temperatures each member of the 122 family exhibits a concomitant structural and magnetic transition. Despite their different structural/magnetic transition temperatures---spanning a range greater than 100 K---and chemical compositions, each of the 122 parent compounds displays the same low-temperature structural and magnetic phases. The tetragonal {\it{I4/mmm}} space group stable at room temperature undergoes a distortion that leads to a low-temperature orthorhombic ({\it{Fmmm}} space group) crystal structure, where the basal plane of the orthorhombic unit cell is rotated by 45$^{\circ}$ with respect to that of the tetragonal unit cell.\cite{Rotter2008, Huang2008} At ambient pressure or without doping, spin-density-wave (SDW), antiferromagnetic (AFM) order occurs simultaneous with the tetragonal-orthorhombic structural transition. The AFM state is characterized by a (101) wavevector (note: the magnetic and orthorhombic unit cells are identical), yielding Fe moments directed along the orthorhombic $a$-axis that are antiferromagnetically arranged along $a$ and $c$ (between Fe layers) and ferromagnetically coupled along $b$. In contrast to the wide range of AFM ordering temperatures ($T_N$), the ordered moment within the 122 family varies only slightly between 0.80 and 1.01 $\mu_{B}$.\cite{Goldman2008, Kaneko2008, Su2009} 

With applied pressure or doping, both the structural and AFM transitions are generally suppressed. In the case of Co or K doping in BaFe$_2$As$_2$, the nominally concomitant structural and magnetic transitions separate from one another, with the structural transition preceding the magnetic transition upon cooling.\cite{Ni2008, Chu2009, Pratt2009, Urbano2010} Near the suppression of the AFM state, with either doping or pressure, superconductivity arises with critical temperatures ranging from roughly 9-47 K. \cite{Han2009, Saha2012} A notable exception to this general observation is the lack of superconductivity in CaFe$_2$As$_2$ under highly hydrostatic pressure conditions.\cite{Yu2009} In addition to superconductivity, an isostructural volume collapse to a collapsed tetragonal phase is seen as a function of pressure as well as for a small minority of chemical substitutions.\cite{Saha2012, Goldman2009, Uhoya2010a, Uhoya2010b, Uhoya2011, Mittal2011, Danura2011} The proximity of the superconducting state with both structural and magnetic instabilities has prompted suggestions that the maximum $T_c$ in the 122 family of compounds could be controlled by structural parameters, magnetic interactions, or both.\cite{Ishida2009, Paglione2010, Johnston2010, Lumsden2010, Yildrim2009} Each of these factors could have ramifications on the pairing symmetry of the superconducting state itself,\cite{Mazin2009} and, as such, exploring the relationships between superconductivity, magnetism, and structural instabilities is an important component of understanding the unconventional, high-temperature superconductivity seen in the ferropnictide compounds.

In this article we report a pressure-dependent structural and electrical transport study of (Ca$_{0.67}$Sr$_{0.33}$)Fe$_2$As$_2$. The isoelectronic substitution of Sr for Ca in this pseudobinary alloy expands the ambient-pressure lattice volume and rapidly increases $T_N$ close to that of SrFe$_2$As$_2$ with with the addition of approximately 30\% Sr.\cite{Kirshenbaum2012} The effects of the larger volume and the enhanced $T_N$ are to expand the phase space occupied by the AFM state to higher temperatures and higher pressures relative to that of pure CaFe$_2$As$_2$, thus pushing the destruction of magnetism to higher pressures and allowing for a larger region of study under pressure.

\section{Experimental Details}
Single crystals of (Ca$_{0.67}$Sr$_{0.33}$)Fe$_2$As$_2$ were synthesized with a flux-growth technique previously described.\cite{Saha2009} The samples were verified with x-ray diffraction to crystallize in the {\it{I4/mmm}} ThCr$_2$Si$_2$-type crystal structure with ambient-pressure lattice constants $a$=3.907 \AA{} and $c$=11.988 \AA. 

Pressure-dependent electrical transport measurements were performed using two pressure cells: (i) a hydrostatic clamp cell employing n-pentane:isoamyl alcohol as a pressure-transmitting medium was used up to 1.1 GPa; and (ii) a designer diamond anvil cell (DAC) loaded with quasihydrostatic solid steatite as a pressure-transmitting medium was used for pressures above 1.76 GPa. The designer DAC was composed of a 300-$\mu$m culet, 8-probe designer diamond anvil\cite{Weir2000, Patterson2000, Jackson2006} paired with a matching standard diamond anvil. In order to facilitate electrical contact with the sample, tungsten contact pads were lithographically deposited onto the microprobes exposed at the culet of the designer diamond anvil.  A non-magnetic MP35N gasket was pre-indented to a thickness of 40~$\mu$m and a 130-$\mu$m hole was drilled in the center of the indentation by means of an electric discharge machine (EDM).  A small, thin crystallite (approximately 70 $\times$ 70 $\times$ x 20~$\mu$m) was placed on the culet of the designer diamond anvil in contact with the tungsten contact pads.  The pressure was calibrated using the shift in the R1 fluorescence line of ruby.\cite{Mao1986, Vos1991}  The ruby R2 fluorescence line remained distinguishable from the R1 line to pressures in excess of 7 GPa, implying a nearly hydrostatic environment below 7 GPa. Temperature-dependent, electrical resistance measurements were performed in a commercial cryostat.

For x-ray diffraction measurements, the DAC was composed of a pair of opposed diamond anvils with 700-$\mu$m culets and a nickel gasket. The gasket was pre-indented to a thickness of 65 $\mu$m and a 250-$\mu$m hole was drilled in the center of the indentation with an EDM.  The (Ca$_{0.67}$Sr$_{0.33}$)Fe$_2$As$_2$ crystals were crushed in a mortar and pestle and loaded into the sample space along with a few small ruby chips for initial pressure calibration and fine Cu powder (3-6 $\mu$m, Alfa Aesar) for {\it{in situ}} x-ray pressure calibration.  A 4:1 methanol:ethanol mixture served as the pressure-transmitting medium.

Room-temperature, angle-dispersive x-ray diffraction (ADXD) experiments were performed at the HPCAT beamline \mbox{16 BM-D} of the Advanced Photon Source at Argonne National Laboratory.  A 5x10 $\mu$m, 29.2 keV (${\lambda}_{inc}$=0.4246 \AA) incident x-ray beam, calibrated with CeO$_2$, was used.  The diffracted x-rays were detected with a Mar345 image plate; exposure times ranged from 300-600~seconds.  2D diffraction patterns were collapsed to 1D intensity versus 2$\Theta$ plots using the program FIT2D.\cite{Hammersley1996}  Pressure-dependent lattice parameters were extracted by indexing the positions of the Bragg reflections using the EXPGUI/GSAS package.\cite{Larson1994, Toby2001}

\section{Results}
\subsection{Crystal structure}\label{CrystalStructure}
\begin{figure}[t]
%COMMENT LINE h=here, t=top, b=bottom, p=separate figure page
\begin{center}\leavevmode
\includegraphics[scale=0.42]{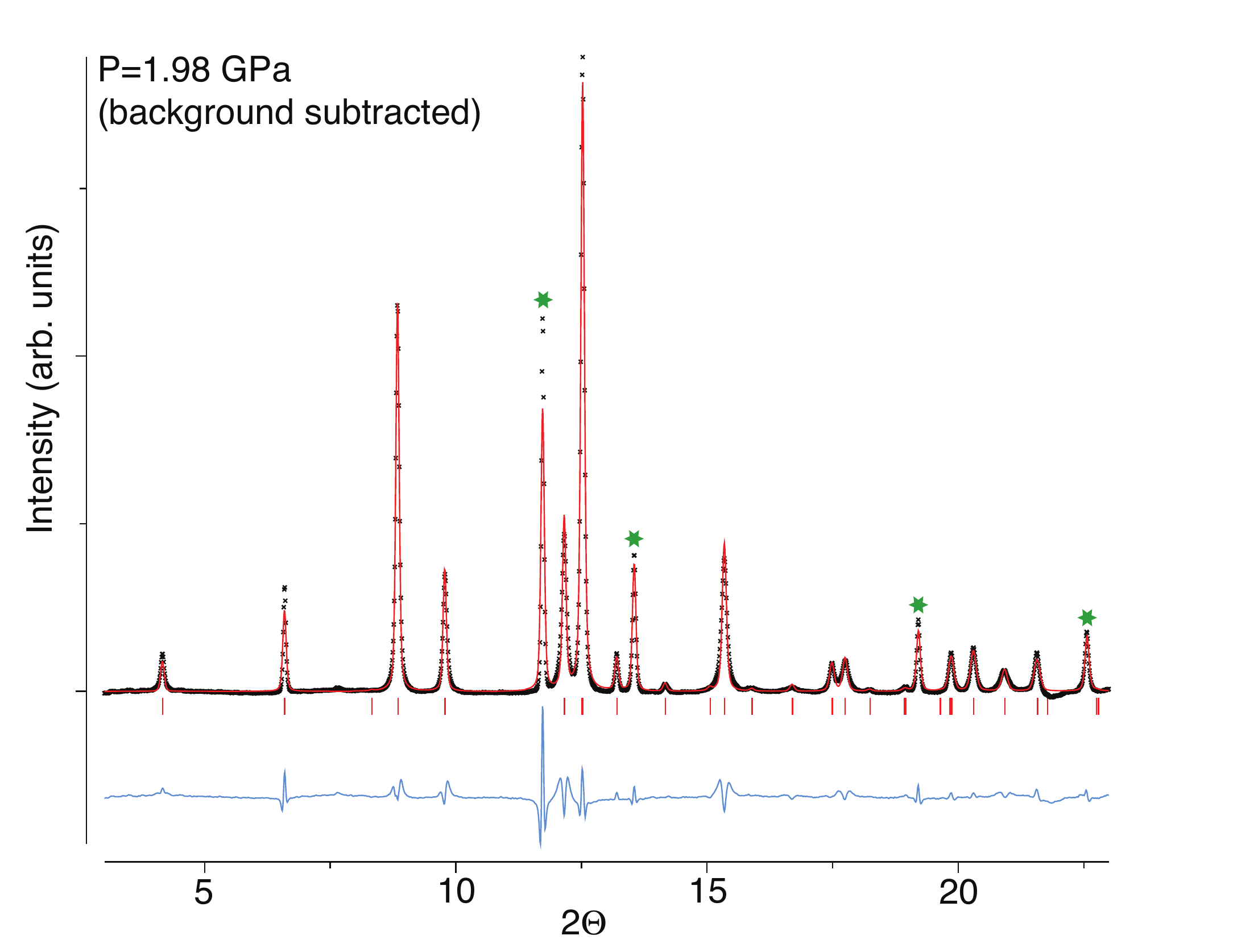}
\caption{(color online) An example x-ray diffraction pattern acquired at 1.98 GPa in a DAC. The refinement runs through the data points as a red line, and the residual of the refinement is shown below the pattern as a light blue line. Bragg reflections of the (Ca$_{0.67}$Sr$_{0.33}$)Fe$_2$As$_2$ sample are shown as red tickmarks, while Bragg peaks from the Cu pressure marker are indicated by the green stars.}\label{XRD}
\end{center}
\end{figure}

A typical, powder x-ray diffraction pattern for (Ca$_{0.67}$Sr$_{0.33}$)Fe$_2$As$_2$, taken at a pressure of 1.98 GPa in the DAC, is shown in Fig.~\ref{XRD}. The Bragg reflections corresponding to the tetragonal {\it{I4/mmm}} structure of (Ca$_{0.67}$Sr$_{0.33}$)Fe$_2$As$_2$ are indicated by the red tickmarks below the data (thin, black crosses), for which the background has been subtracted.The green stars represent the positions of the Bragg peaks of the Cu pressure marker. The diffraction pattern is well described, as indicated by the absence of additional peaks in the pattern, by including only a combination of (Ca$_{0.67}$Sr$_{0.33}$)Fe$_2$As$_2$ and Cu. The (Ca$_{0.67}$Sr$_{0.33}$)Fe$_2$As$_2$ specimen displays a preferred orientation, with the crystallites of the powder tending to form small platelets aligned with the $c$-axis parallel to surface of the culet of the diamond anvil ({\it{i.e.}}, parallel to the incident x-ray beam). This preferred orientation results in a relative decrease in the intensity of the (00l) reflections and an increase in the intensity of the (hk0) reflections. Nonetheless, a full refinement (red line through data) of the diffraction pattern results in a good fit to the data, allowing for determination of the lattice parameters as well as the $z$-coordinate of the As atoms.

%%%%%%%%%%%%%%%%%%%%%%%%%%%%%%%%%%%
%%%%%%%%%%%%%%%%%%%%%%%%%%%%%%%%%%%

\begin{figure}[t]
%COMMENT LINE h=here, t=top, b=bottom, p=separate figure page
\begin{center}\leavevmode
\includegraphics[scale=0.4]{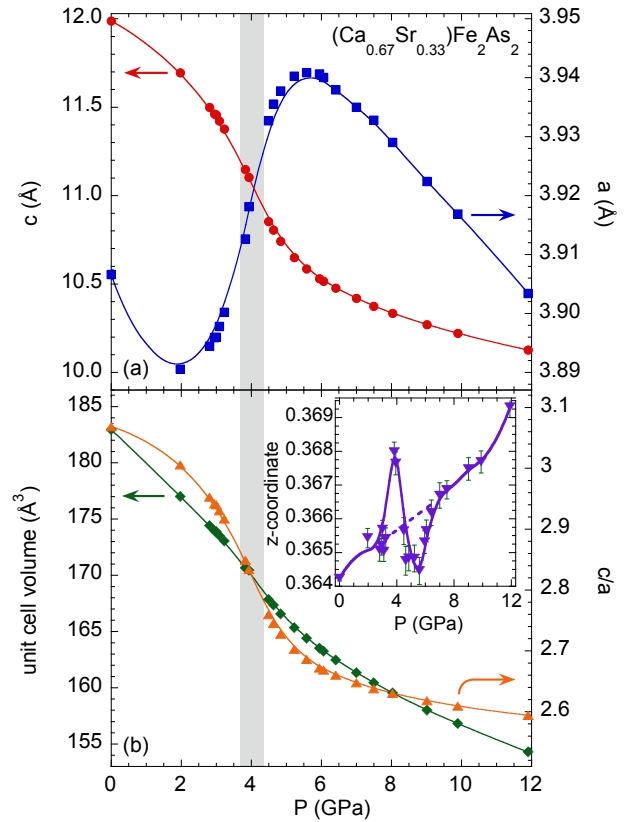}
\caption{(color online) Structural parameters extracted from refinements of x-ray diffraction patterns under pressure: (a) tetragonal lattice parameters $c$ (red circles, left axis) and $a$ (blue squares, right axis), (b) unit cell volume (green diamonds, left axis) and $c/a$ ratio (orange triangles, right axis). The inset shows the refined z-coordinate of the As site as a function of pressure. In all cases, lines are guides to the eye.}\label{Lattice}
\end{center}
\end{figure}

The structural parameters extracted from refinements of the x-ray diffraction data under pressure are shown in Fig.~\ref{Lattice} up to 12 GPa. Ambient-pressure values are from [\onlinecite{Saha2011}]. With increasing pressure, the $c$-axis of the tetragonal unit cell monotonically decreases, but with a steeper slope between roughly 2 and 6 GPa. The $a$-axis, on the other hand, increases with pressure within the same 2-6 GPa range, followed by a more conventional compression for pressures in excess of 6 GPa. The unit cell volume and the $c/a$ ratio (Fig.~\ref{Lattice}b) naturally reflect the pressure dependences of the lattice parameters, with both quantities exhibiting an increased slope centered around 4 GPa. 

%%%%%%%%%%%%%%%%%%%%%%%%%%%%%%%%%%%
%%%%%%%%%%%%%%%%%%%%%%%%%%%%%%%%%%%

These pressure-dependent evolution of the structural parameters shown in Fig.~\ref{Lattice} indicate the presence of an isostructural volume collapse, identical to that seen in the other pure, alkaline-earth 122 compounds: CaFe$_2$As$_2$, SrFe$_2$As$_2$, and BaFe$_2$As$_2$. The structural parameters all exhibit inflection points near 4 GPa, providing a consistent estimate for the volume-collapse transition pressure (vertical, grey bar in Fig.~\ref{Lattice}) in (Ca$_{0.67}$Sr$_{0.33}$)Fe$_2$As$_2$. Above 4 GPa, (Ca$_{0.67}$Sr$_{0.33}$)Fe$_2$As$_2$ is in the collapsed tetragonal phase. The $z$-coordinate of the As atoms, a free parameter within the ThCr$_2$Si$_2$ structure, also exhibits an anomaly near the volume-collapse transition as seen in the inset of Fig.~\ref{Lattice}b. The $z$-coordinate increases slightly at low pressures before exhibiting a significant increase near 4 GPa. Above 4 GPa, the $z$-coordinate decreases before increasing and recovering toward the general trend (dashed line) seen at low pressure, suggesting a correlation between the As atoms and onset of the collapsed tetragonal phase.

\subsection{Electrical transport}\label{ElectricalTransport}
\begin{figure}[t]
%COMMENT LINE h=here, t=top, b=bottom, p=separate figure page
\begin{center}\leavevmode
\includegraphics[scale=0.3]{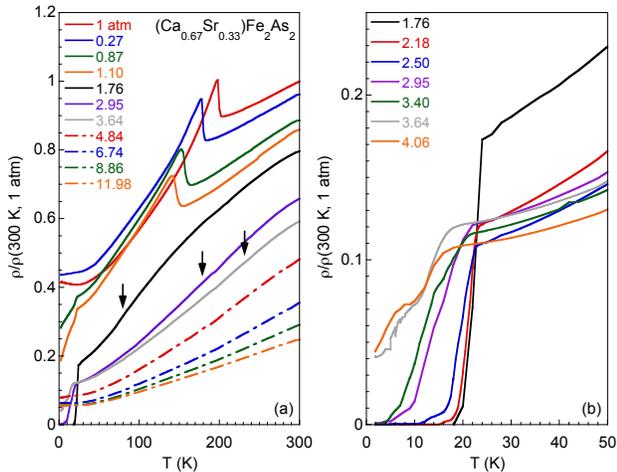}
\caption{(color online) Normalized electrical resistivity as a function of temperature for selected pressures (denoted in GPa unless otherwise specified). The magnetic/structural transition and the onset of superconductivity are visible in (a). The downward arrows ($T_{ct}$) indicate inflection points in the electrical resistivity curves (see text). The evolution of the superconducting transition with pressure is highlighted in (b).}\label{Rho}
\end{center}
\end{figure}

The electrical resistivity $\rho$ as a function of temperature for selected pressures is presented in Fig.~\ref{Rho}. The electrical resistivity data have been normalized such that the ambient pressure value of $\rho$(300 K) is equal to one. In the ambient-pressure curve, the concomitant magnetic and structural transition is evident as a pronounced jump near 200 K. With applied pressure, $\rho$(300 K) decreases and the magnetic/structural transition is smoothly suppressed, disappearing between 1.10 and 1.76 GPa. There is no evidence suggesting a splitting of the structural and magnetic transitions. Within the magnetic state at 0.87 and 1.10 GPa, there is a rapid reduction in resistivity just below 20 K; this behavior is reminiscent of the strain-induced superconductivity observed in pure SrFe$_2$As$_2$ crystals,\cite{Saha2009} although a full resistive transition is lacking here within the magnetic state of (Ca$_{0.67}$Sr$_{0.33}$)Fe$_2$As$_2$.

%%%%%%%%%%%%%%%%%%%%%%%%%%%%%%%%%%%
%%%%%%%%%%%%%%%%%%%%%%%%%%%%%%%%%%%

At 1.76 GPa, the lowest measured pressure where the signature of the magnetic/structural transition is no longer visible, the electrical resistivity displays an inflection point (downward arrows in Fig.~\ref{Rho}a) near 80 K and full resistive superconducting transition at $T_c$=22.2 K. Higher pressures reduce $T_c$, as clearly seen in Fig.~\ref{Rho}b, but substantially increase the temperature of the inflection point. Previous experiments on CaFe$_2$As$_2$ revealed a similar occurrence and pressure-dependent behavior of this inflection point.\cite{Torikachvili2008a}

\begin{figure}[t]
%COMMENT LINE h=here, t=top, b=bottom, p=separate figure page
\begin{center}\leavevmode
\includegraphics[scale=0.3]{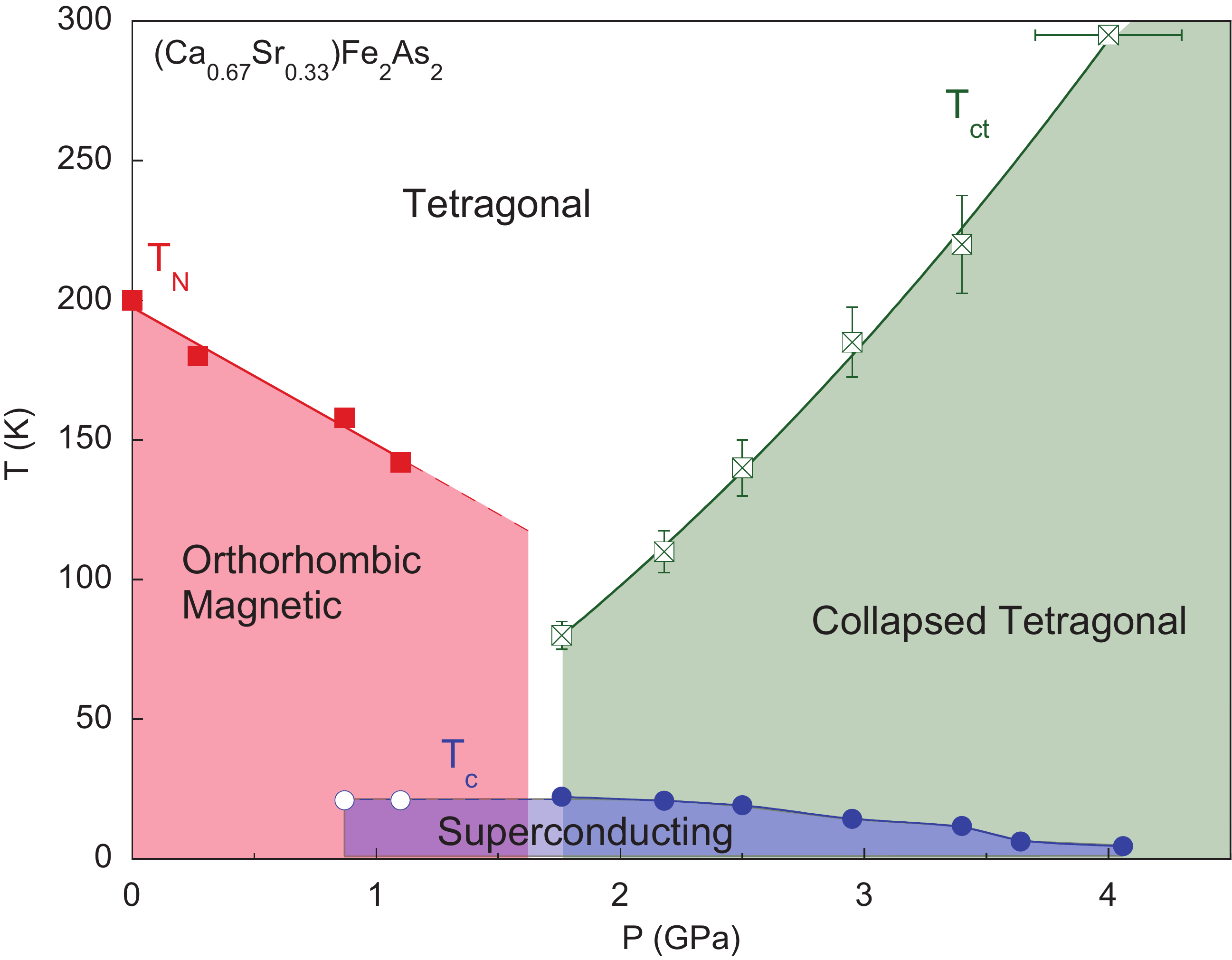}
\caption{(color online) Phase diagram of (Ca$_{0.67}$Sr$_{0.33}$)Fe$_2$As$_2$ showing the suppression of magnetism ($T_N$ - red squares), the development of superconductivity ($T_c$ - blue circles), and the progression of the volume collapse transition ($T_{ct}$ - green, crossed squares) with pressure. The room-temperature value of $T_{ct}$ is determined from x-ray diffraction data, all other data points are from electrical transport measurements. The open, blue circles at lower pressures represent incomplete superconducting transitions. Lines and shaded regions are guides to the eye.}\label{Diagram}
\end{center}
\end{figure}

The features extracted from the pressure-dependent electrical resistivity measurements are collected in the phase diagram of Fig.~\ref{Diagram}. The closed, red squares represent the onset of magnetic order ($T_N$) and its associated structural transition, while the closed, blue circles reveal the evolution of $T_c$ with pressure. The open, blue circles at 0.87 and 1.10 GPa indicate incomplete transitions possibly associated with strain-induced filamentary superconductivity. The inflection point, seen for $P>$1.76 GPa, is shown as open, green squares. The pressure dependence of the inflection point in the electrical resistivity intersects with the volume collapse transition pressure at room temperature (described in \ref{CrystalStructure}), leading to the conclusion that the inflection point seen in the temperature-dependent electrical resistivity signifies the onset of the collapsed tetragonal phase, a conclusion consistent with previous pressure-dependent and Rh-substitution studies of CaFe$_2$As$_2$\cite{Torikachvili2008a, Torikachvili2008b, Yu2009, Canfield2009, Danura2011} We thus denote this feature ({\it{i.e.}}, the inflection point in the electrical resistivity) as $T_{ct}$. Bulk superconductivity, as inferred from a complete resistive transition, is seemingly limited to the collapsed tetragonal phase.

%%%%%%%%%%%%%%%%%%%%%%%%%%%%%%%%%%%
%%%%%%%%%%%%%%%%%%%%%%%%%%%%%%%%%%%
\section{Discussion}\label{Disc}
Given the strong link between the appearances of superconductivity and the collapsed tetragonal phase in the 122 ferropnictide family of superconductors, it is naturally important to explore what driving mechanisms or correlations may be responsible for each phenomena.

\subsection{Isostructural volume collapse}
At room temperature, the isostructural volume collapse in (Ca$_{0.67}$Sr$_{0.33}$)Fe$_2$As$_2$ occurs near 4 GPa (\ref{CrystalStructure}); the volume collapse transition shifts to lower pressures with reduced temperature (\ref{ElectricalTransport}). The evolution of the position of the As atoms within the unit cell (given by the $z$-coordinate in Fig.~\ref{Lattice}) upon passing through the volume collapse transition suggests that the As atoms are involved in this transition. 

%%%%%%%%%%%%%%%%%%%%%%%%%%%%%%%%%%%
\begin{figure}[t]
%COMMENT LINE h=here, t=top, b=bottom, p=separate figure page
\begin{center}\leavevmode
\includegraphics[scale=0.34]{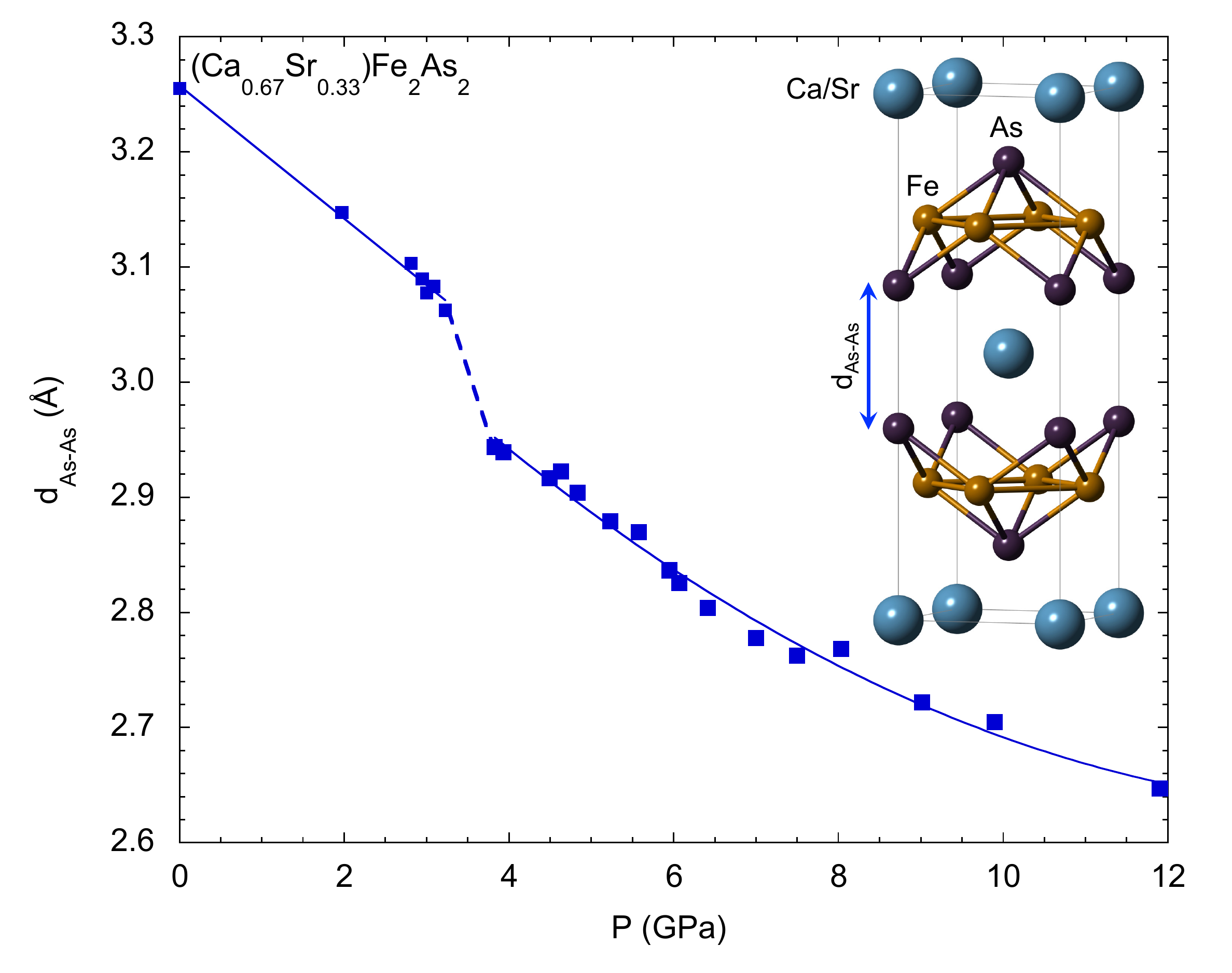}
\caption{(color online) Pressure dependence of the As-As distance ($d_{As-As}$) across the mirror plane of the crystal structure shown in the inset. Lines through the data are guides to the eye.}\label{As_Spacing}
\end{center}
\end{figure}
%%%%%%%%%%%%%%%%%%%%%%%%%%%%%%%%%%%

Figure~\ref{As_Spacing} shows the interlayer As-As spacing across the mirror plane of the unit cell, $d_{As-As}$, as a function of pressure at room temperature. From ambient pressure, the As-As spacing decreases continuously, and nearly linearly, with applied pressure, reaching a value of $d_{As-As}$=3.06 \AA{} at 3.2 GPa. Between 3.2 and 3.8 GPa, $d_{As-As}$ abruptly decreases to a value of 2.94 \AA{}, a 4\% reduction occurring over 0.6 GPa. Further pressure causes a continuous, monotonic decrease in $d_{As-As}$ up to the highest pressure measured. The onset of the collapsed tetragonal phase in (Ca$_{0.67}$Sr$_{0.33}$)Fe$_2$As$_2$, therefore, is signified by a collapse in the As-As separation across the mirror plane of the unit cell. From Fig.~\ref{As_Spacing}, the midpoint of this collapse occurs when $d_{As-As}$=3.0~\AA. 

\begin{figure}[t]
%COMMENT LINE h=here, t=top, b=bottom, p=separate figure page
\begin{center}\leavevmode
\includegraphics[scale=0.35]{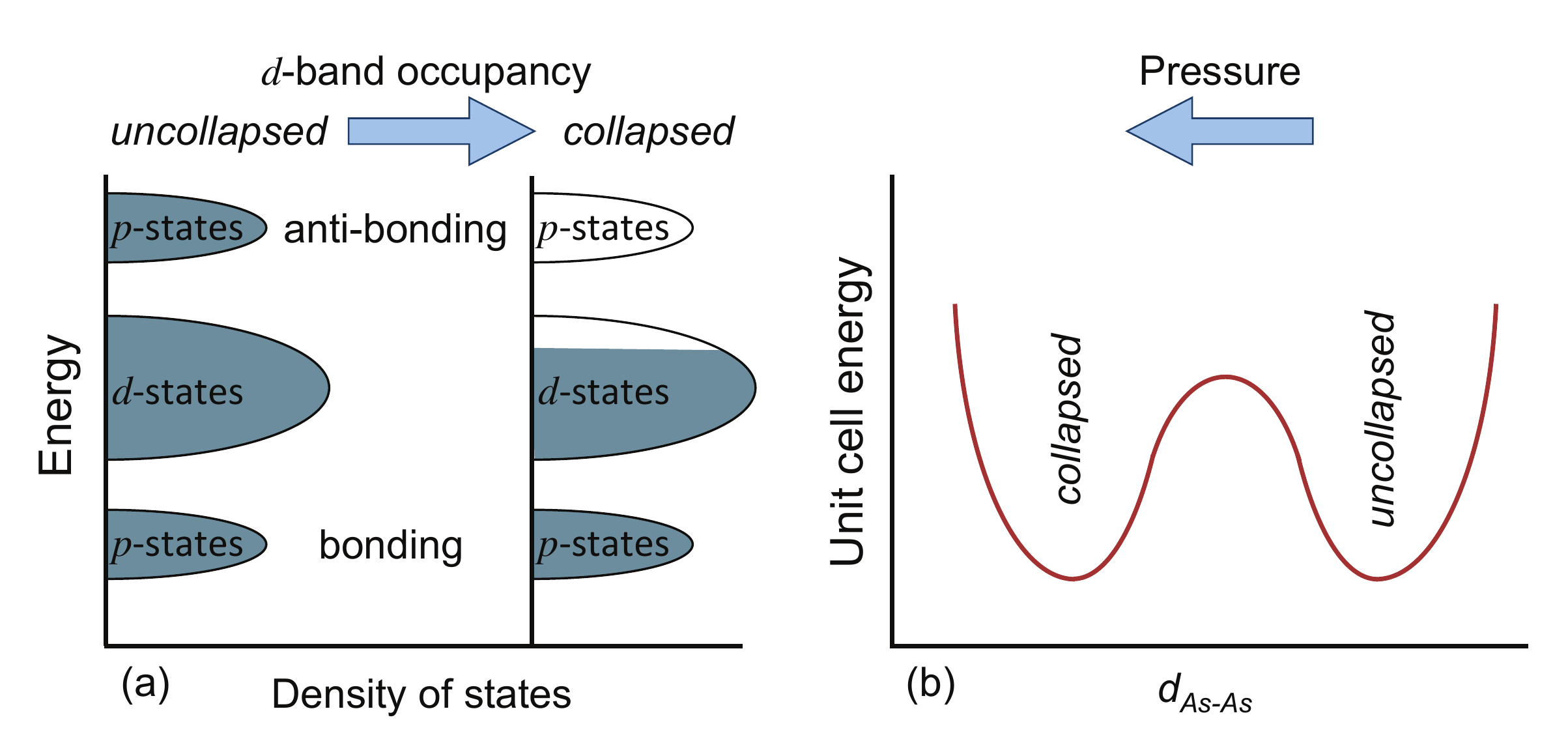}
\caption{(color online) (a) Schematic density of states of the B$_2$As$_{2}^{-2}$ layers for a hypothetical AB$_2$As$_2$ compound. The Fermi level lowers with increasing {\it{d}}-band occupancy, depleting the As-As anti-bonding states, creating an As-As mirror plane bond, and collapsing the structure. (b) Schematic unit cell energy (adapted from [\onlinecite{Hoffman1985}]) versus $d_{As-As}$ for a hypothetical AB$_2$As$_2$ specimen; pressure or {\it{d}}-electron element substitution should push $d_{As-As}$ leftward, providing a driving force for the collapsed phase.}\label{DOS}
\end{center}
\end{figure}

The tendency for a collapse across the mirror plane of the ThCr$_2$Si$_2$-type structure has been discussed previously. Hoffman and Zheng formulated this collapse for BaMn$_2$P$_2$,\cite{Hoffman1985} but the effect can be generalized to other compounds with this structure. For the purpose of discussion we refer to a general formula AB$_2$X$_2$. The basic description of the collapse put forth by Hoffman and Zheng is predicated on the chemistry of the B$_2$X$_{2}^{-2}$ layer, which yields a schematic density of states (Fig.~\ref{DOS}a) with X-X bonding and anti-bonding $p$-states separated by the $d$-states arising from the B atom. As the atomic number of B increases within a row of the Periodic Table, the Fermi level shifts downward, leaving the anti-bonding $p$-states of the density of states unfilled, resulting in the development of an X-X bond across the mirror plane of the structure. X-X bonding across the width of the unit cell does not occur because that dimension is fixed by the $a$ lattice parameter, which is at least partly set by the size of the A cation and typically larger than 3.5~\AA.\cite{Just1996} There is thus a chemical route to describing the uncollapsed and collapsed tetragonal phases of the ThCr$_2$Si$_2$-type structure, which sheds light on the mechanism under pressure. 

%%%%%%%%%%%%%%%%%%%%%%%%%%%%%%%%%%%
%%%%%%%%%%%%%%%%%%%%%%%%%%%%%%%%%%%

Full electronic structure calculations by Hoffman and Zheng indicate that for BaMn$_2$P$_2$ there is a maximum structural energy (Fig.~\ref{DOS}b) when the P-P distance is about 2.7~\AA.\cite{Hoffman1985} In BaMn$_2$P$_2$, the P-P bond length is thus shifted roughly 0.5~\AA{} above the bare P-P bond length.\cite{Pyykko2009} If a similar value of the X-X bond-length shift occurs in (Ca$_{0.67}$Sr$_{0.33}$)Fe$_2$As$_2$, then the As-As bond length would shift from the bare As-As bond length of 2.4~\AA{} to about 2.9~\AA, in excellent agreement with the value $d_{As-As}$=3.0~\AA{} defining the volume collapse transition pressure.

The onset of such As-As bonding would naturally be directed along the $c$-axis of the unit cell (inset, Fig.~\ref{As_Spacing}), and would tend to pull the previously weakly connected FeAs cages toward one another, accounting for the contraction of the crystallographic $c$-axis upon entering the collapsed tetragonal phase. By conservation, the increase in bonding between mirror plane As atoms would likely reduce the Fe-As bond strength, which would relax the FeAs cages, alter the Fe-As bond angles, and increase the $a$-axis of the unit cell as seen experimentally. 

In addition to the structural consequences of the development of this new As-As bond within the structure, there are likely electronic structure effects. Band structure calculations conclude that the transition into the collapsed tetragonal phase results in a downward shift of the bands relative to the uncollapsed phase and a reduction in the density of states at the Fermi level.\cite{Yildrim2009, Goldman2009} Indeed, magnetotransport measurements in rare-earth doped CaFe$_2$As$_2$ show a dramatic reduction in the magnitude of the Hall coefficient upon cooling through the volume collapse transition.\cite{Saha2012} In addition, first-principles calculations for CaFe$_2$As$_2$ show that the strength of the Fe-As bonds, the As-As mirror plane bonds, and the Fe spin-state, and henceforth magnetism, are strongly coupled.\cite{Yildrim2009} Thus, the disappearance of magnetic order with the onset of the collapsed tetragonal phase may be an unsurprising consequence of the electronic structure mandated by the collapsed tetragonal phase. 

\begin{figure}[t]
%COMMENT LINE h=here, t=top, b=bottom, p=separate figure page
\begin{center}\leavevmode
\includegraphics[scale=0.30]{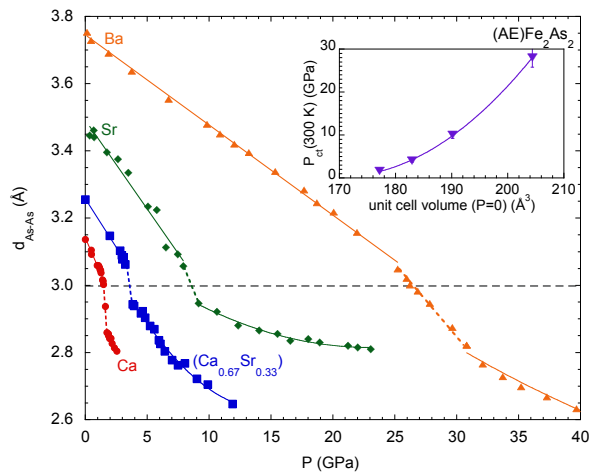}
\caption{(color online) Comparison of the pressure dependence of $d_{As-As}$ for the alkaline earth (AE)Fe$_2$As$_2$ compounds. The horizontal dashed line indicates $d_{As-As}=$3.0 \AA. Data for CaFe$_2$As$_2$, SrFe$_2$As$_2$, and BaFe$_2$As$_2$ are from references [\onlinecite{Goldman2009}], [\onlinecite{Uhoya2011}], and [\onlinecite{Mittal2011}], respectively. Inset: volume collapse transition pressure, $P_{ct}$, as a function of unit cell volume at $P$=0. Lines through data points are guides to the eye.}\label{As_Compare}
\end{center}
\end{figure}

The onset of As-As interlayer bonding, as indicated by the contraction of $d_{As-As}$, is not unique to (Ca$_{0.67}$Sr$_{0.33}$)Fe$_2$As$_2$. In fact, the other pure alkaline earth 122 ferropnictide superconductors as well as some of their rare-earth-doped counterparts\cite{Saha2012} display identical behavior in $d_{As-As}$, albeit at different pressures. Figure~\ref{As_Compare} shows $d_{As-As}$ as a function of pressure for members of the (AE)Fe$_2$As$_2$ family. The horizontal line represents the onset of As-As bonding at $d_{As-As}$=3.0~\AA. Each compound has been shown to undergo the volume collapse transition, and, accordingly, each displays an abrupt contraction of the As-As separation. As the size of the alkaline earth element increases, the unit cell volume of the crystal structure increases, and $d_{As-As}$ increases. A natural consequence of this unit cell volume expansion is that a larger pressure is required to achieve sufficient lattice compression to invoke As-As bonding across the mirror plane of the unit cell, and the volume collapse transition concordantly shifts to higher pressures with increasing atomic radius of the alkaline earth element (inset, Fig.~\ref{As_Compare}).

%%%%%%%%%%%%%%%%%%%%%%%%%%%%%%%%%%%
%%%%%%%%%%%%%%%%%%%%%%%%%%%%%%%%%%%

\subsection{As-Fe-As bond angles}
\begin{figure}[t]
%COMMENT LINE h=here, t=top, b=bottom, p=separate figure page
\begin{center}\leavevmode
\includegraphics[scale=0.34]{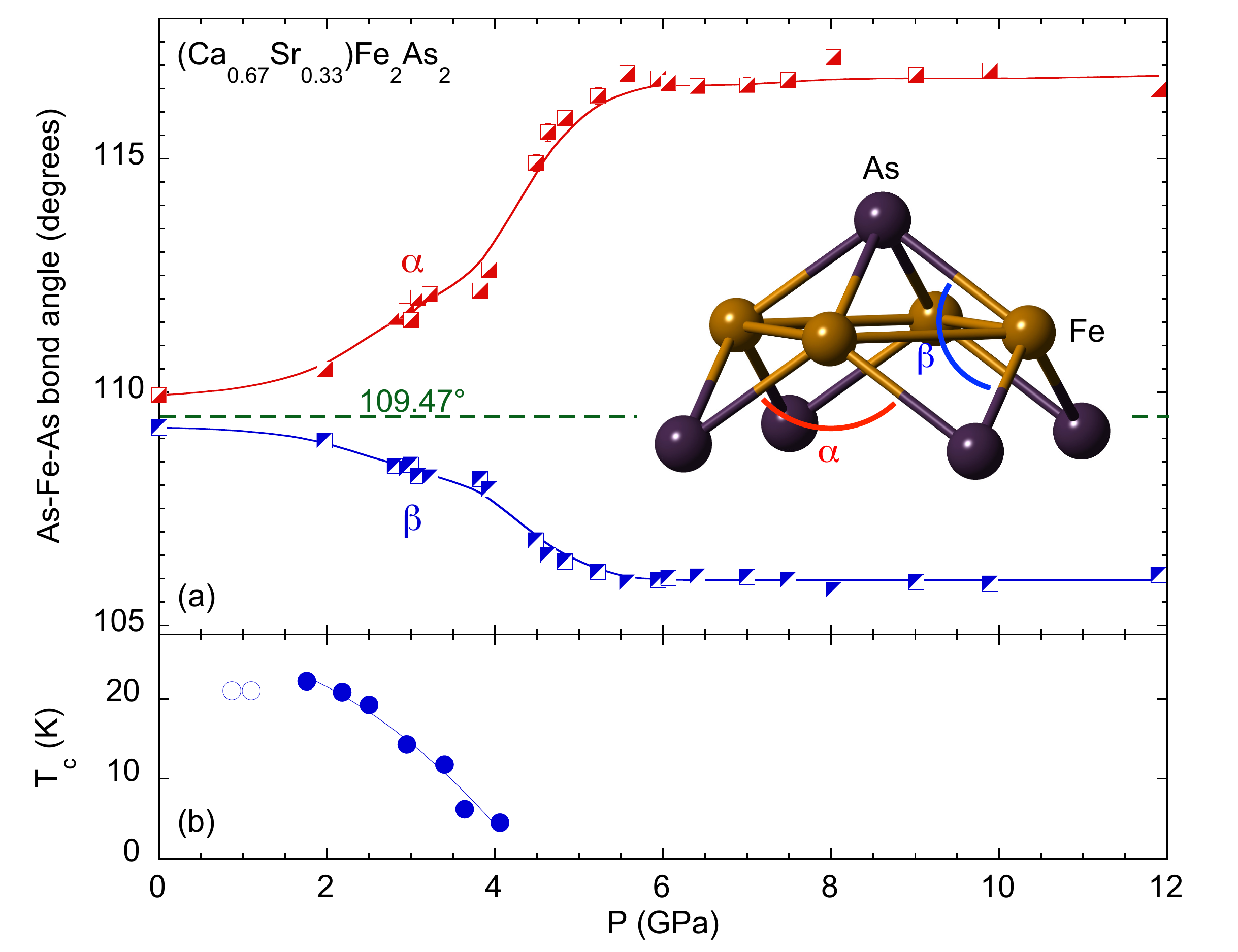}
\caption{(color online) (a) The room-temperature values of the two-fold, $\alpha$, and four-fold, $\beta$, As-Fe-As bond angles as a function of pressure. The ideal tetrahedral angle (109.47$^{\circ}$) is marked by the horizontal, green, dashed line. The inset defines $\alpha$ and $\beta$ within the corrugated FeAs component of the crystal structure. (b) The pressure dependence of $T_c$ on the same pressure axis; symbols identical to Fig.~\ref{Diagram}. Lines are guides to the eye.}\label{BA}
\end{center}
\end{figure}

Early studies of the iron-bearing oxypnictide superconductors noted an empirical correlation between the As-Fe-As bond angles and the maximum observed $T_c$.\cite{Ishida2009, Lee2008} Within the corrugated FeAs layers of the crystal structure, there are two As-Fe-As bond angles: the two-fold or intralayer angle, denoted as $\alpha$; and the four-fold or interlayer angle, denoted as $\beta$ (see the inset of Fig.~\ref{BA}).\cite{Johnston2010} Due to the crystal structure, these two bond angles move oppositely (as $\alpha$ increases, $\beta$ decreases), but if the As atoms surrounding the Fe atoms are perfectly tetrahedrally coordinated, then $\alpha$=$\beta$=109.47$^{\circ}$. In reference [\onlinecite{Lee2008}], it was found that as the As-Fe-As bond angles of LnFeAsO$_{1-y}$ approached this ideal tetrahedral angle, $T_c$ reached a maximum value near 55~K. Since then, this general empirical relationship has been noted in nearly all families of ferropnictide superconductors.\cite{Paglione2010, Johnston2010} However, like many ``rules'' in condensed-matter physics, there are exceptions, notably CsFe$_2$As$_2$ with a low $T_c$=2.6 K and As-Fe-As bond angles of 109.58$^{\circ}$ and 109.38$^{\circ}$.\cite{Gooch2010}

%%%%%%%%%%%%%%%%%%%%%%%%%%%%%%%%%%%
%%%%%%%%%%%%%%%%%%%%%%%%%%%%%%%%%%%

Figure~\ref{BA}a displays the pressure dependence of the $\alpha$ and $\beta$ As-Fe-As bond angles at room temperature. At ambient pressure, the corrugated FeAs layers exhibit coordination close to the ideal tetrahedral configuration ($\alpha$=109.93$^{\circ}$, $\beta$=109.24$^{\circ}$). Applied pressure drives the structure away from this ideal tetrahedral coordination, and, above the volume collapse transition, the bond angles settle into relatively pressure-independent values significantly disparate from the ideal tetrahedral condition. 

The evolution of $T_c$ with pressure is reproduced in Fig.~\ref{BA}b for comparison with the bond-angle evolution. While $T_c$ seems to be correlated with the bond angles, with $T_c$ decreasing as the bond angles deviate from that of the ideal tetrahedron, it should be emphasized that a one-to-one correspondence is likely too simple of an explanation. The bond angle data in Fig.~\ref{BA}a was determined at room temperature, while the determination of $T_c$ is clearly at low temperature. From the phase diagram in Fig.~\ref{Diagram}, it can be seen that the superconducting phase occurs within the collapsed tetragonal phase. If the relatively constant bond angles seen in the collapsed tetragonal phase at room temperature are representative of that phase even at low temperatures, then one might expect that the maximum in $T_c$ would occur with bond angles $\alpha{\approx}$116$^{\circ}$ and $\beta{\approx}$106$^{\circ}$, distinctly deviating from the ideal tetrahedral angle. Unfortunately, no low-temperature structural data were acquired in this study, but low-temperature structural characterization of pure and rare-earth doped CaFe$_2$As$_2$, which still exhibit superconductivity, indicate that the bond angles tend {\it{away}} from the ideal tetrahedral angle upon cooling.\cite{Kreyssig2008, Saha2012} Furthermore, the phase diagram of Fig.~\ref{Diagram} reveals that the superconducting state occurs in proximity to the destruction of magnetic order, its associated structural transition, and the occurrence of an isostructural volume collapse, certainly suggesting that the appearance of superconductivity may be correlated with factors other than structural parameters at room temperature. 

\subsection{Structural and electronic phase diagrams}
\begin{figure}[t]
%COMMENT LINE h=here, t=top, b=bottom, p=separate figure page
\begin{center}\leavevmode
\includegraphics[scale=0.31]{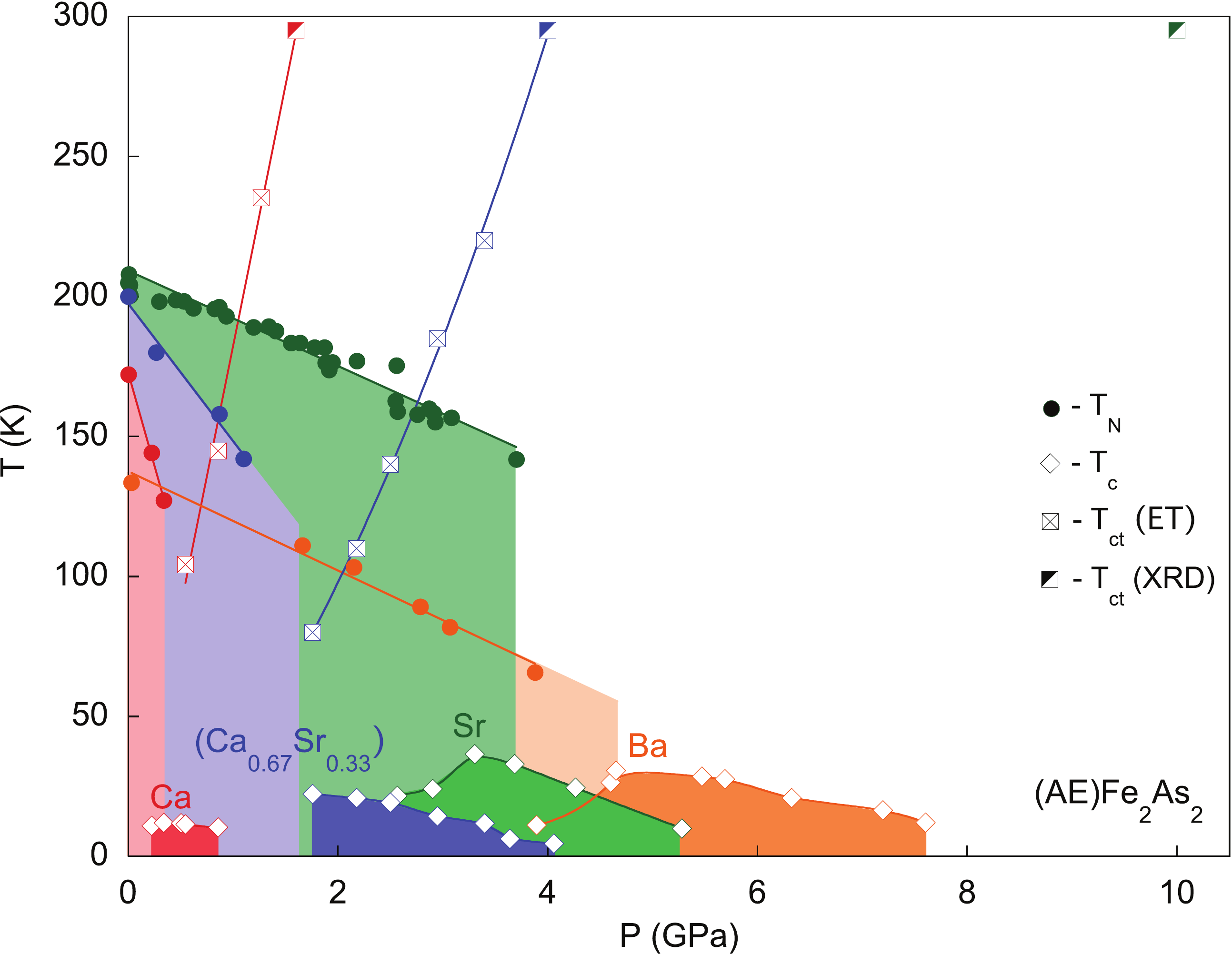}
\caption{(color online) Comparison of the temperature-pressure phase diagrams for (AE)Fe$_2$As$_2$. Closed circles represent $T_N$, open diamonds represent $T_c$, and crossed or diagonal boxes represent $T_{ct}$. The volume collapse transition as determined by electrical transport and x-ray diffraction are denoted by ET and XRD, respectively. Electrical transport data for CaFe$_2$As$_2$, SrFe$_2$As$_2$, and BaFe$_2$As$_2$ are from [\onlinecite{Torikachvili2008a, Torikachvili2008b, Colombier2009}]; structural data are from [\onlinecite{Goldman2009, Uhoya2011, Mittal2011}]. Lines and shaded regions are guides to the eye. }\label{Phase_Compare}
\end{center}
\end{figure}

The structural and electronic phase diagram of (Ca$_{0.67}$Sr$_{0.33}$)Fe$_2$As$_2$ interpolates very well with the phase diagrams of its end member compounds as well as the related BaFe$_2$As$_2$ compound. These phase diagram are shown together in Fig.~\ref{Phase_Compare}, highlighting the qualitative similarities within the AEFe$_2$As$_2$ system. Each compound obeys some general behavioral rules. At ambient-pressure, each compound undergoes a structural/magnetic transition ($T_N$) at sub-ambient temperatures. $T_N$ is suppressed with applied pressure and abruptly disappears well above $T=0$. Superconductivity develops around this discontinuous destruction of magnetism and persists as a several-GPa-wide dome or half-dome in P-T space with a maximum $T_c$ occurring close to the pressure at which $T_N$ abruptly vanishes. Each compound exhibits a pressure-induced volume collapse at room temperature, and the volume collapse transition $T_{ct}$ occurs with positive slope in P-T space ({\it{i.e.,}} $dT_c/dP>0$) where it has been measured. The pressure axis of Fig.~\ref{Phase_Compare} does not extend far enough to include the volume collapse transition in BaFe$_2$As$_2$, which, from the data shown in Fig.~\ref{As_Compare}, occurs near 26 GPa.\cite{Mittal2011}

In CaFe$_2$As$_2$ and (Ca$_{0.67}$Sr$_{0.33}$)Fe$_2$As$_2$, measurements of the $T_{ct}(P)$ line strongly suggest that the volume collapse itself is likely responsible for the abrupt destruction of magnetic order. This further implies that magnetism is limited to the orthorhombic phase, and that the collapsed tetragonal phase does not support magnetic order. While the destruction of magnetism may be linked to the onset of the collapsed tetragonal phase, whether that destruction is driven by a reduction in the Fe moments, an altering of some exchange coupling, or a more subtle change in the electronic structure is an open question likely requiring both theoretical and experimental input to reach a conclusion.

%%%%%%%%%%%%%%%%%%%%%%%%%%%%%%%%%%%
%%%%%%%%%%%%%%%%%%%%%%%%%%%%%%%%%%%

Unlike the volume collapse transition, the facts about the occurrence of superconductivity in the AEFe$_2$As$_2$ systems are less clear. With the use of more hydrostatic pressure conditions, researchers have found that the superconducting dome, as defined by complete resistive transitions or susceptibility data, in the parent compounds is generally excluded from the magnetic, orthorhombic phase.\cite{Alireza2009, Colombier2009} The results on (Ca$_{0.67}$Sr$_{0.33}$)Fe$_2$As$_2$ are consistent with this finding even with the less hydrostatic steatite pressure-transmitting medium used in this study. However, it is imperative to note that a study using He as the pressure-transmitting medium revealed the {\it{absence}} of superconductivity in both the orthorhombic (magnetic) {\it{and}} the collapsed tetragonal phases of CaFe$_2$As$_2$.\cite{Yu2009} A simple, quantitative shift in the pressure at which superconductivity appears as a function of hydrostaticity would not be entirely surprising, but the qualitative difference ({\it{i.e.}}, a complete lack of superconductivity) as a function of pressure media creates a conundrum regarding the appearance of superconductivity in the AEFe$_2$As$_2$ systems.

While the generic phase diagrams of chemical substitution and doping look strikingly similar, there are subtle differences that question the roles of structural and magnetic instabilities and suggest that a one-to-one correspondence between pressure and substitution is too simple. Studies of electron- and hole-doped BaFe$_2$As$_2$ have revealed a splitting of the nominally coupled paramagnetic-antiferromagnetic and tetragonal-orthorhombic phase transitions with increasing dopant content.\cite{Ni2008, Chu2009, Pratt2009, Urbano2010} Superconductivity is seen to develop within the antiferromagnetic, orthorhombic phase with both Co and K substitution in BaFe$_2$As$_2$,\cite{Ni2008} but the optimal $T_c$ is achieved within the paramagnetic, uncollapsed tetragonal phase. Though superconductivity develops within the antiferromagnetic state, the suppression of that state, at least to some degree, seems to be a necessary ingredient for superconductivity. Unlike doping of the BaFe$_2$As$_2$ end member, Rh-doping into CaFe$_2$As$_2$ does not reveal a significant splitting of the paramagnetic-antiferromagnetic and tetragonal-orthorhombic phase transitions, but $T_N$ is nonetheless suppressed with increasing doping.\cite{Danura2011} Superconductivity is seen only in the paramagnetic, uncollapsed tetragonal phase, and the onset of a doping-induced collapsed tetragonal phase destroys superconductivity. Recently, a doping study using rare earth (RE) elements in Ca$_{1-x}$RE$_{x}$Fe$_2$As$_2$ has shown that the components of electron doping and chemical pressure can be effectively separated, and that each of these components plays a different role in manifesting superconductivity.\cite{Saha2012} Within the Ca$_{1-x}$RE$_{x}$Fe$_2$As$_2$ series, superconductivity occurs in either the uncollapsed or collapsed tetragonal phases, but not within the antiferromagnetic, orthorhombic phase. 

Assimilating the doping- and pressure-dependent phase diagrams of the 122 systems in order to better understand the nature of the high-temperature superconductivity seen therein is a challenging problem. Given the overwhelming evidence that electron- or hole-doping plays an important role in the development of superconductivity within the chemically substituted AEFe$_2$As$_2$ systems and given the occurrence of strain-induced superconductivity in SrFe$_2$As$_2$,\cite{Saha2009} it is tempting to posit that the small superconducting dome seen under pressure may be a product of some effective doping induced by non-hydrostatic pressure conditions, possibly resulting not only from the pressure-transmitting medium but from the volume collapse transition itself. Alternatively, without high-fidelity, low-temperature structural data under pressure, it is difficult to exclude structural phase inhomogeneity ({\it{e.g.}}, coexistence of the collapsed and uncollapsed tetragonal phases) as a possible cause of the proximity of superconductivity to the observed structural instabilities. More work demarcating the $T_{ct}(P)$ lines in SrFe$_2$As$_2$ and BaFe$_2$As$_2$ under pressure may help to illuminate any possible connections between pressure-induced superconductivity and the isostructural volume collapse.

\section{Conclusions}
The compound (Ca$_{0.67}$Sr$_{0.33}$)Fe$_2$As$_2$ under pressure behaves intermediate between its two end member parent compounds. With applied pressure, the concomitant structural/magnetic transition is suppressed with no evidence favoring a splitting of the two nominally coupled transitions. The AFM, orthorhombic phase is abruptly cut off by an isostructural volume collapse resulting in a collapsed tetragonal phase. The volume collapse transition is driven by the development of As-As bonding across the mirror plane of the crystal structure, contracting the $c$-axis of the unit cell and likely affecting the Fe-As bonding and potentially the magnetic state of the Fe atoms. The collapsed tetragonal phase supports superconductivity with a maximum $T_c$ near 22 K. There is no obvious structural parameter that defines the magnitude of $T_c$, but the proximity of the superconducting phase to the suppression of magnetism as well as the onset of the collapsed tetragonal phase suggests that magnetic interactions and/or structural inhomogeneity may both play a role in the development of pressure-induced superconductivity in these systems.

\section{Acknowledgments}
We are grateful to Z. Jenei and K. Visbeck for assistance with cell preparations.  JRJ and STW are supported by the Science Campaign at Lawrence Livermore National Laboratory.  Portions of this work were performed under LDRD (11-LW-003).  Lawrence Livermore National Laboratory is operated by Lawrence Livermore National Security, LLC, for the U.S. Department of Energy, National Nuclear Security Administration under Contract DE-AC52-07NA27344.  Portions of this work were performed at HPCAT (Sector 16), Advanced Photon Source (APS), Argonne National Laboratory. HPCAT is supported by CIW, CDAC, UNLV and LLNL through funding from~DOE-NNSA, DOE-BES and NSF. Use of the Advanced Photon Source, an Office of Science User Facility operated for the U.S. Department of Energy (DOE) Office of Science by Argonne National Laboratory, was supported by the U.S. DOE under Contract No. DE-AC02-06CH11357. Beamtime was provided through the Carnegie-DOE Alliance Center (CDAC). This work was partially supported by AFOSR-MURI Grant No. FA9550-09-1-0603.  YKV acknowledges support from DOE-NNSA Grant No. DE-FG52-10NA29660.

\end{document}